\documentclass[prl,aps,twocolumn,floats,showpacspsfig]{revtex4}
\usepackage{amssymb}

\usepackage{epsfig}

\newcommand{\be}{\begin{equation}}
\newcommand{\ee}{\end{equation}}
\newcommand{\bea}{\begin{eqnarray}}
\newcommand{\eea}{\end{eqnarray}}

\newcommand{\p}{\partial}
\newcommand{\s}{\sigma}

\newcommand{\la}{\langle}
\newcommand{\ra}{\rangle}
\newcommand{\rd}{\mbox{d}}
\newcommand{\ri}{\mbox{i}}
\newcommand{\re}{\mbox{e}}

\begin{document}
%\draft
\title{Striped pnictides as new strongly correlated systems.}

\author{ 
%A. V. Chubukov and
 A.  M. Tsvelik}
\affiliation
%\institute
{ Department of  Condensed Matter Physics and Material Science, Brookhaven National Laboratory, Upton, NY 11973-5000, USA}

\date{\today}

\begin{abstract}
%\abstract{
The recent experimental evidence in favor of stripe ordering in  ferropnictides  [Chuang {\it et. al.}, Science {\bf 327},181 (2010)] points to the new field of study for strongly correlated systems. Here we argue that due to the smallness of the chemical potential in ferropnictides the effects of stripe modulation are likely to be much more pronounced than in the cuprates leading to stronger degree of one-dimensionality. This may allow to treat the striped pnictides as an array of coupled one-dimensional spin liquids making them  a good testing ground for various ideas concerning   strongly correlated states originally proposed for the cuprates. 
\end{abstract}

%\pacs{71.10.Pm} {Fermions in reduced dimensions}
%\pacs{72.80.Sk} {Insulators }

\maketitle
%\narrowtext

\sloppy
%\section{Introduction}
%\section{Striped pnictides? }

  One of the ideas grown from the attempts to understand the fascinating properties of copper oxides is the conviction that doping of Mott insulators reveals a latent 
  tendency for superconducting pairing already hidden in the parent system. Whether this idea is relevant to the cuprates remains controversial. What we have at the moment is a theoretical demonstration of its validity for one-dimensional models such as models of ladders \cite{fisher}. It has been rigorously demonstrated that if the undoped ladder is a Mott insulator, then under doping it gives rise to a superconducting quasi long range order. For 3D array of ladders (LA) this would give rise to full fledged superconductivity. Naturally, it would be interesting to find or to custom made materials fitting the model description of LAs to check the validity of the theory. One example of a real material considered as close approximation of  the LA ideal is the so-called "telephone number" compound Sr$_{14-x}$Ca$_x$ Cu$_{24}$ O$_{41}$. Although this material possesses a lot of interesting properties, it appears to be too complicated due to the fact that its structure contains not just CuO ladders, but also chains \cite{girsh}.  
  
  In this paper we suggest that one can search for realizations of LA not just among CuO-based systems, such as the telephone number compound, but among nearly compensated metals such as ferropnictides. The Fermi surface of such metals consists of small electron and hole pockets so that  the Fermi energy is much smaller than the 
  bandwidth $ \epsilon_F << W$. Systems of that kind can be considered as strongly correlated in the sense that, as was demonstrated
 %AC
in Ref.\cite{chubukov08},
% by Chubukov \cite{chub},
 the interactions undergo strong renormalization and at energies close to $\epsilon_F$ the effective Hamiltonian is substantially different from the one given by the band structure theory. This feature rises the red flag for numerical studies of finite systems meaning  that size effects in these systems must be really severe. Fortuitously it turns out that instead of complicating matters, the renormalization simplifies the interaction pattern. Namely, interactions in the broad conduction and valence bands drive the system towards higher symmetry simultaneously increasing their own strength. The resulting effective theory in the region of energies $\sim \epsilon_F$ is described by a highly symmetric Hamiltonian where instabilities in several different channels (Spin Density Wave, superconductivity and 
%AC suggest
Charge Density Wave)
% etc.) 
compete with each other
 %~\cite{kim}.
% Kim et al considered only a symmetric case
 It is believed that finite $\epsilon_F$  breaks the symmetry and favors some particular type of order 
%AC \cite{kim}. This was discussed in my review
\cite{chub}.
 There is, however, a possibility that at zero doping no order is chosen, like in spin liquid or  band insulator (after all in the pnictides there are two nonequivalent sites in the elementary cell). Such possibility is easily realized for the undoped 1D ladder which then becomes superconducting under doping \cite{fisher}.
  
  We suggest that a "custom made" material for a doped spin liquid would be a "striped" ferropnictide, that is ferropnictide subject to a one-dimensional periodic potential of a moderate amplitude $\epsilon_F < U << W$ (further in the text we provide more detailed criteria). One can imagine such potential being produced when a pnictide film is grown on a suitable substrate or occur naturally. The recently performed experiments indicate that the latter possibility is realized  in the ferropnictide Ca(Fe$_{1-x}$Co$_{x}$)$_2$ As$_2$ 
  %, FeSe$_{1-x}$Te$_{x}$ \cite{ar},\cite{tranq1} 
  which  naturally develops a strong in-plane anisotropy (stripe ordering ?) and become rather one-dimensional as a result \cite{tiang}. Theoretically one-dimensional version of the pnictide Hamiltonian  has recently been considered by Berg {\it et. al.}\cite{berg}. These authors applied DMRG method to the four-chain lattice Hamiltonian.  Here we assume a much smaller degree of one-dimensionality which occurs for a moderate  potential modulation not affecting the states with energies larger than $U << W$. This allows us to assume that the renormalization process has taken its toll  and for energies smaller than $\epsilon_F$ one can use the simplified Hamiltonian with enlarged symmetry. The advantages of our model are  numerous. First, we have a  one-dimensional model where non-perturbative methods can be applied. Second, the situation we consider has a good chance to describe reality. Third, the stripe modulation may enhance the pairing strength, as it probably does for the cuprates \cite{tranquada},\cite{2DHub}.  Fourth,  by considering a symmetric model we simplify the discussion. 
  
  The corresponding Hamiltonian density has the form considered in \cite{kim}:
  \bea
 &&  {\cal H} = - c^+_{\s}\p_x^2 c_{\s} + f^+_{\s}\p_x^2f_{\s} - u_0(n_c -n_f)^2 + \label{kim}
\\ 
 && u_2\Big(c^+_{\uparrow}c^+_{\downarrow}f_{\downarrow}f_{\uparrow} + h.c.\Big)  - k^2_F(n_c - n_f) - \mu(n_c + n_d) \nonumber
  \eea
  where $ \mu$ measures  a deviation from a perfect nesting.
%b) I would suggest to put minus in front  of u_0 term here for consistency with earlier works, c) there is also c^+f f^+c interaction. It does not flow to zero under RG but becomes relatively small compared to u_0 and u_2. Nice thing is that this interaction renormalizes via itself [x^\dot = (...)*x], so you may simply say that you set its bare value to zero, and it does not apper under RG
It is worth noticing that the Coulomb interaction, $(n_c + n_f)^2$ term, renormalizes to zero \cite{AAC}. It is also possible to show that the disparity between electron and hole masses only affects the coefficients in two-dimensional RG equations without changing the low energy Hamiltonian (\ref{kim}). 

 Assuming that the interactions are weak we linearize the spectrum close to the Fermi points:
  \bea
  c = R\re^{-\ri k_F x} + L\re^{\ri k_F x}, ~~ f = r\re^{\ri k_F x} + l\re^{-\ri k_F x},
  \eea
  where $k_F$ is the Fermi momentum at $\mu =0$. 
%AC You set k_F the same for c and f fermions. Not true if \mu \neq 0
  The kinetic energy acquires the standard form (we neglect the difference in the Fermi velocities of electrons and holes):
  \bea
  T = \ri v(- R^+_{\s}\p_x R_{\s} + L^+_{\s}\p_x L_{\s}) + \ri v(- r^+_{\s}\p_x r_{\s} + l^+_{\s}\p_x l_{\s})
  \eea
  The Umklapp interaction becomes
  \bea
  && V_2 = u_2\Big\{\Big[L_{\uparrow}L_{\downarrow}r^+_{\downarrow}r^+_{\uparrow} + R_{\uparrow}R_{\downarrow}l^+_{\downarrow}l^+_{\uparrow} + \nonumber\\
  && \Big(R_{\uparrow}L_{\downarrow} + L_{\uparrow}R_{\downarrow} \Big)\Big(l^+_{\downarrow}r^+_{\uparrow} + r^+_{\downarrow}l^+_{\uparrow} \Big)\Big] + h.c.\Big\} \label{V2}
  \eea
  The densities are
  \bea
  && n_c = (R^+_{\s}R_{\s} + L^+_{\s}L_{\s}) + \re^{-2\ri k_F x}L^+_{\s}R_{\s} + \re^{2\ri k_F x}R^+_{\s}L_{\s}\nonumber\\
  && n_f = (r^+_{\s}r_{\s} + l^+_{\s}l_{\s}) + \re^{-2k_F\ri x}r^+_{\s}l_{\s} + \re^{2\ri k_F x}l^+_{\s}r_{\s}
  \eea
  The bosonization rules are standard:
  \bea
  L_{\s} = \frac{\xi^{(1)}_{\s}}{\sqrt{\pi a_0}}\re^{\ri\sqrt{4\pi}\varphi^{(1)}_{\s}}, ~~ R_{\s} = \frac{\xi^{(1)}_{\s}}{\sqrt{\pi a_0}}\re^{-\ri\sqrt{4\pi}\bar\varphi^{(1)}_{\s}} \label{bos}
  \eea
%AC Should it be \bar\varphi^{(1)}_{\s}} ?
  and the same for $r,l$ with index 1 being replaced with 2. Here $\varphi,\bar\varphi$ are chiral bosonic fields and $\xi$ are coordinate independent Klein factors:
  \bea
  \{ \xi_{\s}^a,\xi_{\s'}^b\} = \delta_{\s\s'}\delta_{ab}
  \eea
  The Klein factors constitute the Clifford algebra and are Dirac $\gamma$-matrices of the O(6) group. 
  
  Substituting (\ref{bos}) into (\ref{V2}) and introducing new fields
  \bea
 &&  \phi_c^{(+)} = \frac{1}{2}\Big[\phi_{\uparrow}^{(1)} + \phi_{\downarrow}^{(1)} + \phi_{\uparrow}^{(2)} + \phi_{\downarrow}^{(2)} \Big]\nonumber\\
  &&  \phi_c^{(-)} = \frac{1}{2}\Big[\phi_{\uparrow}^{(1)} + \phi_{\downarrow}^{(1)} - \phi_{\uparrow}^{(2)} - \phi_{\downarrow}^{(2)} \Big]\nonumber\\
  &&  \phi_s^{(+)} = \frac{1}{2}\Big[\phi_{\uparrow}^{(1)} - \phi_{\downarrow}^{(1)} + \phi_{\uparrow}^{(2)} - \phi_{\downarrow}^{(2)} \Big]\nonumber\\
  &&  \phi_s^{(-)} = \frac{1}{2}\Big[\phi_{\uparrow}^{(1)} - \phi_{\downarrow}^{(1)} - \phi_{\uparrow}^{(2)} + \phi_{\downarrow}^{(2)} \Big]
  \eea
 we get
  \bea
 &&  V = 4u_2\frac{\xi_{\uparrow}^{(1)}\xi_{\downarrow}^{(1)}\xi_{\downarrow}^{(2)}\xi_{\uparrow}^{(2)}}{(\pi a_0)^2}\times\\
  && \cos[\sqrt{4\pi}\theta_c^{(-)}]\Big\{\cos[\sqrt{4\pi}\phi_c^{(+)}] + \cos[\sqrt{4\pi}\phi_s^{(+)}] + \cos[\sqrt{4\pi}\phi_s^{(-)}] \Big\},\nonumber
\eea
where $\phi = \varphi + \bar\varphi, ~~ \theta = \varphi - \bar\varphi$. 
This interaction can be refermionized:
\bea
V = u_2\bar\psi_0\psi_0\sum_{a=1}^3(\bar\psi_a\psi_a)
\eea
 where $\bar\psi_a = (L^+,R^+)_a, \psi^T_a = (R,L)_a$.
%AC You defined 4 components later, suggest to move here
 The new fermions transform in the vector representation of SO(8) group and should not be confused with original ones. 
 The density-density interaction gives rise to the term
  \bea
  V_0 = - u_0\sum_{a\neq b}(\bar\psi_a\psi_a)(\bar\psi_b\psi_b)
  \eea
  As a result we get the Gross-Neveu model with $Z_2\times$O(6) symmetry:
  \bea
  && {\cal H} = \ri v\sum_{a=1}^3(- R^+_a\p_x R_a + L^+_a\p_xL_a) + \nonumber\\
  && \ri v (- R^+_0\p_x R_0 + L^+_0\p_xL_0) + \nonumber\\
  && -u_0\sum_{a\neq b}(R^+_aL_a + L^+_aR_a)(R^+_bL_b + L^+_bR_b) - \nonumber\\
  && u_0' (R^+_aL_a + L^+_aR_a)^2 +  \nonumber\\
  && u_2(R^+_0L_0 + L^+_0R_0)\sum_{a=1}^3(R^+_aL_a + L^+_aR_a)  \nonumber\\
&& - \mu(R_1^+R_1 + L^+_1L_1) -h(R^+_2R_2 + L^+_2L_2) \label{GN}
\eea
(here the index 0 corresponds to $(c,-)$, 1 to $(c,+)$, 2 and 3 to $(s,\pm)$). The RG equations can be extracted from \cite{fisher}:
\bea
&& \dot u_0 = - 4u_0^2 - 2u_2^2, ~~ \dot u_2 = -(u_0' + 5u_0)u_2, \nonumber\\
&& \dot u_0' = - 4u_0^2 - 2u_2^2 \label{RG}
\eea
%AC A few points. a) as I write earlier, suggest to reverse the sign of u_0 from the beginning, b) you may wish to put comparison in 2D. In the same notations as yours, as I understand, in 2D case u_0 = u'_0, u^\dot_0 = - 2 u^2_0 -2 u^2_2, u^\dot_2 = -6 u_0 u_2. c) the bare values of  u_0 and u'_0 are presumably the same
%AC For my education: in 2D I had two couplings (u_0 and u_2), and, apparently, no third coupling u'_0 is generated by RG. Am I missing smth in 2D, or 1D is special. Would be nice if put a phrase on this. 

 Let us consider the undoped ($\mu =0$) case first.  Analysis of Eqs.(\ref{RG}) shows that the interactions scale to strong coupling under rather general conditions. At strong coupling the RG trajectories asymptotically restore the SO(8) symmetry $u_0 = u_0' = |u_2|$. The spectrum is gapful and at the SO(8) symmetric point is known exactly (one can find a detailed discussion in \cite{fisher}). The fact that the spectrum is gapful is an indication that the undoped state is not  a superconductor. In the ground state fields $\phi_c^{(+)},\phi_s^{(\pm)}$ freeze at $\Phi =0$, the field $\theta_c^{(-)}$ freezes at $0$ or $\sqrt{\pi}/2$ depending on sign of $u_2$
%AC in 2D, the sign of u_2 differentiates between s^++ and s^{+-} pairing, AND
% also between SDW and spin currents. Is there an analogy to this in 1D (I mean, the analogy to SDW vs spin currents)
There is no {\it local} order parameter corresponding to these field configurations. This situation is known as $d$-Mott insulator \cite{fisher}. It is interesting that the coherent spin excitations (vector particles) are  emitted at the Neel wave vector connecting the centers of electron and hole pockets. The corresponding operator is the staggered spin current:
\bea
&& {\bf N} = \ri [c^+\vec\s f - f^+\vec\s c],\label{N}\\
&& N^+ \approx \frac{4\ri\xi^{(1)}_{\uparrow}\xi^{(2)}_{\downarrow}}{\pi a_0}\times\nonumber\\
&& \cos[\sqrt{\pi}(\phi_c^{(+)} + \theta_c^{(-)})]\cos[\sqrt{\pi}(\phi_s^{(-)} + \theta_s^{(+)})]\nonumber
\eea

 At finite doping the chemical potential always exceeds the charge gap.
%AC I suggest to add a phrase about what happens with RG eqs.
 Then field $\phi_c^{(+)}$ becomes massless, but all other fields remain massive with doping dependent diminished 
 spectral gaps. The correlation functions in this case have been calculated in \cite{konik}. The quasi long range 
%AC suggest 
superconducting
 order emerges with the order parameters
 \bea
 && \Delta_c = c_{\uparrow}c_{\downarrow} \approx L_{\uparrow}R_{\downarrow} + R_{\uparrow}L_{\downarrow} = \nonumber\\
 && \frac{2\xi^{(1)}_{\uparrow}\xi^{(1)}_{\downarrow} }{(\pi a_0)}\re^{\ri\sqrt{\pi}[\theta_c^{(+)} + \theta_c^{(-)}]}\cos\{\sqrt\pi[\phi_s^{(+)} + \phi_s^{(-)}]\}\label{SC1}\\
 && \Delta_f = f_{\uparrow}f_{\downarrow} \approx l_{\uparrow}r_{\downarrow} + r_{\uparrow}l_{\downarrow} = \nonumber\\
 && \frac{2\xi^{(2)}_{\uparrow}\xi^{(2)}_{\downarrow} }{(\pi a_0)}\re^{\ri\sqrt{\pi}[\theta_c^{(+)} - \theta_c^{(-)}]}\cos\{\sqrt\pi[\phi_s^{(+)} - \phi_s^{(-)}]\} \label{SC2}
 \eea
 The relation between the signs of the amplitudes is determined by the sign of $u_2$ (minus for $u_2 >0$). The phase of the OPs is $\theta_c^{(+)}$. The scaling dimensions at are $d =1/4K$, where $K$ is the Luttinger liquid parameter in the charge sector. In the regime where the forward scattering is weak $K \approx 1$. The pairing susceptibility is strongly divergent: 
 \be
 \chi_P \sim T^{-2 +2d} = T^{-3/2}.
 \ee
and hence the resulting superconductivity  is of a strongly non-BCS nature.

 Now we would like to some aspects of the stripe formation. To simplify the calculations we  model the  stripe as a periodic potential made of a sum of parabolic ones: 
 \bea
 && U(y) = \sum_{n=-\infty}^{\infty} u(y-nb), \\
 && u(x) = U_0[2\pi y/b]^2\theta(|y|-b/2).\nonumber
 \eea
 In the tight binding approximation the lowest band has the wave functions 
 \bea
 && \psi(q,y) = \frac{1}{\sqrt{\pi \xi N}}\sum_n\exp[-(y-nb)^2/2\xi^2]\re^{\ri qn}, \nonumber\\
 && \xi^{-2} = (2\pi/b)\sqrt{U_0m} \label{psi}
\eea
($N$ being the number of stripes) with the transverse dispersion 
\bea
&&\epsilon(q) = 2t\cos (qb), \nonumber\\
&&  t = \frac{\pi}{2}\sqrt{U_0/mb^2}\exp\Big[-(\pi/2)\sqrt{U_0mb^2}\Big]. \label{t}
\eea
For our model calculations to be valid we need this transverse dispersion to be somewhat  smaller that the 1D gaps. The corrections in $(t/M)$ can be taken into account in spirit of \cite{rrts} where a very similar model of 2-leg ladders coupled by weak transverse tunneling was considered. Then with expression for transverse tunneling (\ref{t}) available, we can estimate the transition temperature. From (\ref{SC1},\ref{SC2}) we conclude that the order parameter amplitude is $\sim M/W$ ($M$ being the gap scale in the doped regime). The Josephson coupling between the stripes is $J \sim t^2/M$. The pairing susceptibility is $\chi_P \sim \rho(\epsilon_F)(M/T)^{3/2}$. The criterion for the transition is $J\chi_P \sim 1$ which gives
\bea
T_c \sim M\Big[(t^2/W)\rho(\epsilon_F) M\Big]^{2/3}
\eea
This 3D transition temperature is significantly smaller than the energy gap $M$. In that sense a hypothetic pnictide stripe superconductor will be similar to the underdoped cuprates where the spin gap greatly exceeds the transition temperature. 

 Naturally, wave function (\ref{psi}) enters in  the momentum dependence of various correlation functions through the formfactor. Thus the correlation functions of various densities (such as, for instance (\ref{N}) at low energies will have the dynamical susceptibility displaying 1D dispersion:
 \bea
 && \la\la {\bf N}(-\omega,-{\bf q}){\bf N}(\omega,{\bf q}')\ra\ra = \label{fact}\\
 && g(q_{\perp})g(q'_{\perp})\sum_m\delta(q_{\perp}-q'_{\perp} - 2\pi m/b)D(\omega,q_{\parallel})\nonumber\\
 && g(q) \approx \frac{1}{\pi \xi}\int \rd y \re^{\ri qy} \exp[-y^2/\xi^2] = \re^{- (q\xi/2)^2}\nonumber
\eea
The transverse wave vector at which the intensity drops sharply is 
\be
q_{max} \approx (2\pi/b)[U_0mb^2/\pi^2]^{1/4}
\ee
Due to a rapid exponential dependence of the transverse tunneling amplitude (\ref{t}) it is conceivable to have $q_{max} < 2\pi/b$ and still have $t < M$. We mention this because
 the factorization  (\ref{fact}) in combination with 1D dispersion of gapped magnetic excitations has been observed in FeTe$_{0.6}$Se$_{0.4}$ \cite{tranq1}.  The ARPES data presented in \cite{tranq1} also show that the low-lying quasiparticle excitations are anisotropic which is consistent with the anisotropy of the magnetic excitation spectrum. 
 
 We conclude the paper with a brief discussion of salient features the experimentalists have to look for. Primarily these are spectral gaps; since model  (\ref{GN}) has a complicated spectrum, the gaps are likely to be different in different dynamical response functions. 
 %In our opinion such evidence already exists. Figs. 2ab from \cite{ar} show the spectrum of magnetic excitations in FeTe$_{0.6}$Se$_{0.4}$. Although the authors give it a different interpretation, the spectrum depicted is strikingly similar to a spectrum of magnetic excitations in the spin liquid
% \bea
 %\epsilon(q)= \sqrt{[v(q -Q)]^2 + \Delta^2}
 %\eea
% with $Q =\pi/a$ being the commensurate wave vector and $\Delta = 5-6$meV being the spin gap.  We suggest that the rounding up of the spectrum at small momenta has been misinterpreted as an evidence for incommensurability. Neutron measurements reported in \cite{tranq1} show that the dispersion is indeed quasi-one-dimensional which agrees with the picture presented in this work.  ARPES data presented in \cite{tranq1} show that the low-lying quasiparticle excitations are also anisotropic which is consistent with the anisotropy of the magnetic excitation spectrum. 
 
 Although in this picture we considered a situation when there is one electron and one hole band at the chemical potential, the number can differ for different pnictides. For instance, the STM measurements of the only known striped pnictide Ca(Fe$_{0.97}$Co$_{0.03}$)$_2$ As$_2$ \cite{tiang} found only one hole band (though strongly one-dimensional) and no evidence for electron bands. A possible explanation is that one electron and one hole band got paired producing a gap and leaving one unpaired band behind. The evidence in favor of such explanation comes from the fact that the observed density of states is strongly energy dependent on the scale $|E| < 0.1$eV and exhibits  a mixed metallic and pseudogap-like shape. In any case, a detailed explanation of these experiments is probably premature and is not a purpose of this paper which primary goal is to attract attention to the subject of striped pnictides. 

 I am  grateful to Maxim Khodas, John Tranquada for fruitful discussions and to Andrey Chubukov for reading the manuscipt and making valuable remarks. 
%AC AVC  acknowledges support from the Institute for Strongly Correlated and Complex Systems at BNL.   
This research was  supported  by US DOE, Office of Basic Energy Science as a part of CES. The Center for Emerging Superconductivity is a DOE 
Energy Frontier Research Center. 
%  under contract number DE-AC02 -98 CH 10886. 

\end{document}